\documentclass[a4paper,11pt]{article}
\usepackage{aaskaiid}
\usepackage{orcidlink}

\usepackage{booktabs}

\title{Faraday Tomography with the SKA: A New Era of Cosmic Magnetism Studies}
\ShortTitle{Faraday Tomography with the SKA}

\author[1,2,3,4]{Miguel Carcamo\orcidlink{0000-0003-0564-8167}}
\ShortName{Miguel Carcamo et al\@.} 
\author[5,6]{Anna Scaife\orcidlink{0000-0002-5364-2301}}
\author[7]{Jeroen Stil\orcidlink{0000-0003-2623-2064}}
\author[8,9]{Russ Taylor\orcidlink{0000-0001-9885-0676}}
\author[10,11]{Jennifer L. West\orcidlink{0000-0001-7722-8458}}
\author[12]{Tessa Vernstrom\orcidlink{0000-0001-7093-3875}}

\affiliation[1]{Universidad de Santiago de Chile (USACH), Facultad de Ingeniería, Departamento de Ingeniería Informática, Santiago, Chile}

\affiliation[2]{Data Observatory Foundation, Santiago, Chile}

\affiliation[3]{Millennium Nucleus on Young Exoplanets and their Moons, YEMS, Chile}

\affiliation[4]{Center for Interdisciplinary Research in Astrophysics and Space Exploration (CIRAS), Universidad de Santiago de Chile, Santiago, Chile}

\affiliation[5]{Jodrell Bank Centre for Astrophysics, University of Manchester, Oxford Road, Manchester M13 9PL, UK}

\affiliation[6]{Alan Turing Institute, Euston Road, London, UK}

\affiliation[7]{Department of Physics and Astronomy, The University of Calgary, 2500 University Drive NW, Calgary, Alberta, Canada}

\affiliation[8]{Inter-University Institute for Data Intensive Astronomy, Department of Astronomy, University of Cape Town, Private Bag X3, Rondebosch, 7701, Cape Town, South Africa}

\affiliation[9]{Inter-University Institute for Data Intensive Astronomy, Department of Astronomy, University of Cape Town, Private Bag X3, Rondebosch, 7701, Cape Town, South Africa}

\affiliation[10]{Dominion Radio Astrophysical Observatory, Herzberg Research Centre for Astronomy and Astrophysics, National Research Council
Canada, PO Box 248, Penticton, BC, V2A 6J9, Canada}
\affiliation[11]{
School of Natural Sciences, University of Tasmania, PO Box 807, Sandy Bay, TAS 7006, Australia}

\affiliation[12]{Australia Telescope National Facility, CSIRO, Space and Astronomy, PO Box 1130, Bentley, WA 6102, Australia}

\emailAdd{miguel.carcamo@usach.cl}
\emailAdd{anna.scaife@manchester.ac.uk}
\emailAdd{jeroen.stil@ucalgary.ca}
\emailAdd{russ@idia.ac.za}
\emailAdd{jennifer.west@utas.edu.au}
\emailAdd{tessa.vernstrom@atnf.csiro.au}

\abstract{The Square Kilometre Array (SKA) represents a significant advancement in radio astronomy, enabling detailed study of cosmic magnetism through Faraday Rotation and Faraday Measurement Synthesis. This chapter provides a comprehensive review of these techniques, tracing their development and illustrating their crucial role in investigating magnetic fields across different cosmic environments.

We focus on Array Assembly 4 (AA4), the final deployment stage, featuring 197 dishes (133 SKA, 64 MeerKAT) with 40 km maximum baselines and frequency coverage from 350 MHz to 15.4 GHz (goal: 24 GHz). These capabilities enable high-resolution Faraday tomography of magnetic structures in galaxies, clusters, and the cosmic web.

We also highlight the earlier AA* stage with 144 dishes (80 SKA, 64 MeerKAT). Band 1 (350--1050 MHz) offers the finest Faraday depth resolution among the dish bands (full resolution $\sim$2.5 rad m$^{-2}$) with a maximum observable depth around 174,708 rad m$^{-2}$. Other bands (Band 2, Band 5a, Band 5b) enable multi-scale studies, probing larger structures with lower resolution but higher maximum depths.

We discuss synergies between Faraday Measurement Synthesis and Image Synthesis techniques, showing how they complement each other in reconstructing magnetic field structures. Simulation results demonstrate the SKA's potential for high-resolution observations. We explore possible enhancements beyond the baseline configuration and synergies with other observatories, emphasizing multi-wavelength astronomy.

This chapter aims to provide a thorough account of how the SKA's phased deployment from AA* to AA4 will transform our understanding of cosmic magnetic fields and drive new discoveries in astrophysics.

\textbf{Key points:} (1) Faraday tomography reconstructs magnetic field structure along the line of sight from polarised emission and is a core tool for cosmic magnetism. (2) The SKA will deliver a major step forward in sensitivity, frequency coverage, and angular resolution, enabling denser and more precise RM grids than current facilities. (3) RM grids with the SKA will transform studies of magnetic fields in galaxy clusters, the Galactic ISM, and the large-scale cosmic web.}

\begin{document}
\maketitle

\section{Introduction}

The study of cosmic magnetism represents one of the most fundamental challenges in modern astrophysics. Magnetic fields permeate the Universe, from the smallest scales of stars and planets to the largest structures of galaxy clusters and the cosmic web. Understanding these magnetic fields is crucial for unravelling the physical processes that shape our Universe, from star formation and galaxy evolution to the dynamics of the intergalactic medium. Despite their critical importance in cosmic evolution, key questions regarding the origin, amplification, and evolution of cosmic magnetism remain unanswered from an observational point of view; significant progress has been made in magnetohydrodynamic (MHD) simulations and theoretical understanding \citep{origin-1, GAENSLER20041003, origin-2, origin-3, origin-4, cosmic_magnetism_review}.

Radio astronomy has advanced substantially over the past decade with the commissioning of facilities such as the Very Large Array (VLA), the Atacama Large Millimetre/submillimetre Array (ALMA), and the Low Frequency Array (LOFAR). The VLA in particular established the foundations for polarisation and Faraday rotation studies. Following the upgrade to the Karl G. Jansky Very Large Array in 2012 and the advent of LOFAR, broadband polarisation measurements became routine, and the technique of Rotation Measure (RM) Synthesis \citep{brentjens} came into widespread use, extending the formalism of the Faraday dispersion function introduced by \citet{burn}. These facilities have significantly extended the ability of the community to study the interstellar medium, galaxy evolution, and large-scale cosmic structures.

A central product of such work is the construction of RM grids (systematic surveys of RMs toward many background sources across the sky, which probe magnetic fields in the intervening medium). Current RM grids illustrate the baseline from which the SKA will advance. The NVSS RM catalogue \citep{Taylor-2009} provides roughly 1 RM per square degree over most of the northern sky. The LOFAR Two-metre Sky Survey achieves $\sim$0.43 RMs deg$^{-2}$ with high Faraday-depth precision \citep{10.1093/mnras/stac3820}; the Polarisation Sky Survey of the Universe's Magnetism (POSSUM) on ASKAP is delivering 30--50 RMs deg$^{-2}$ over the southern sky with median RM uncertainties of $\sim$1 rad\,m$^{-2}$ \citep{Vanderwoude2024, GaenslerPOSSUM}; and SPICE-RACS provides $\sim$4--7 RMs deg$^{-2}$ over the southern sky from ASKAP \citep{SPICE-RACS2023}. The SKA will improve on these benchmarks through greater sensitivity (enabling fainter polarised sources and denser grids), wider fractional bandwidth and frequency coverage (improving Faraday depth resolution and RM precision), and higher angular resolution (enabling finer spatial sampling of magnetic structures). With the construction of next-generation radio telescopes, in particular the Square Kilometre Array (SKA) and its precursors, new opportunities arise to probe cosmic magnetism through detailed observations of polarised synchrotron emission and its modification by Faraday rotation; this chapter introduces the role of the SKA in that endeavour.

The SKA, as the world's largest radio telescope, is poised to revolutionise our understanding of cosmic magnetism \citep{Vernstrom01.2026.SKA}. This chapter focuses on how its capabilities, including the phased deployment from Array Assembly * (AA*) to Array Assembly 4 (AA4), will enable studies of magnetic fields across cosmic scales using Faraday rotation and Faraday Measurement Synthesis techniques.

The SKA will allow us to address several fundamental questions about cosmic magnetism:

\begin{itemize}
    \item What is the origin and evolution of magnetic fields in galaxies and galaxy clusters?
    \item How do magnetic fields influence star formation and galaxy evolution?
    \item What role do magnetic fields play in the dynamics of the intergalactic medium?
    \item How do magnetic fields contribute to the formation and evolution of cosmic structures?
    \item What is the nature of magnetic fields in the early Universe?
\end{itemize}

These questions are particularly relevant for studying various cosmic environments, including radio halos and relics in galaxy clusters \citep{halos-3, halos-2, stuardi2021, halos-1, Vacca01.2026.SKA}, cosmic filaments \citep{Carretti2022}, and the magnetic structure of galaxies probed via polarised emission from background radio galaxies \citep{distances-2, distances-1}. The SKA will support detailed mapping of magnetic fields across these environments, providing insights into their role in cosmic evolution.

To address these questions, several science cases will require dense Rotation Measure (RM) grids. Such grids are essential for mapping the two-dimensional projected distribution of magnetic fields (as seen on the plane of the sky) and understanding their role in astrophysical processes. Critically, both Mid and Low RM-grids will be essential, as they provide complementary capabilities: SKA-Mid offers higher angular resolution and sensitivity for detailed studies of magnetic field structures, while SKA-Low provides the finest Faraday depth resolution necessary for probing fine-scale magnetic field features, where the latter is limited by the density of the RM-grid. The combination of both frequency ranges enables comprehensive multi-scale studies of cosmic magnetism that cannot be achieved with either alone.

Faraday rotation, the rotation of the plane of polarisation of electromagnetic waves as they propagate through a magnetised plasma, provides a powerful probe of magnetic fields along the line of sight. When combined with the SKA's wide frequency coverage and sensitivity, this technique allows us to construct detailed maps of magnetic fields in various astrophysical environments. Truly understanding cosmic magnetic fields requires hundreds or thousands of measurements of distant radio galaxies, pulsars, and diffuse sources; next-generation telescopes will enable very large surveys of the sky, opening a new window into the magnetic Universe \citep{Jarvis:20184F, gcluster_meerkat}. The SKA's multi-frequency capabilities, spanning 350 MHz to 15.4 GHz (with a goal of 24 GHz), will provide fine resolution in Faraday depth space.

The construction of the SKA is progressing through multiple deployment stages. The first operational phase will consist of 512 aperture array stations for SKA-Low and 197 parabolic dish antennas for SKA-Mid (133 SKA and 64 MeerKAT). The staged approach will progressively improve angular resolution through several key milestones: AA0.5 (2024--2025, 8 km baselines achieving 4.0'' at 4.6 GHz), AA1 (2025--2026, 20 km baselines reaching 1.6''), AA2 (2026--2027, 120 km baselines providing 0.27''), and finally AA* (2027--2028) and AA4 (2029) with 150 km baselines delivering 0.21'' at 4.6 GHz.

This chapter is structured as follows. We begin by examining these technical capabilities, then explore how they translate into scientific opportunities for studying magnetic fields across different cosmic environments.

A key aspect of our discussion will be the synergy between Faraday Measurement Synthesis and Image Synthesis techniques. We will present simulation results that demonstrate the SKA's potential for high-resolution observations and discuss how these capabilities can be enhanced through synergies with other observatories.

The chapter concludes by looking forward to the transformative impact of the SKA on our understanding of cosmic magnetism and the new discoveries that await us in this new era of radio astronomy.

\section{Faraday Rotation}

\subsection{Introduction to Faraday Rotation}

Faraday Rotation is a phenomenon that occurs when electromagnetic waves pass through a magnetised medium. It is a key tool for studying magnetic fields in the Universe. The phenomenon was first described by Michael Faraday in 1845 and has since become one of the most powerful techniques for probing cosmic magnetic fields. The modern theoretical framework for understanding Faraday rotation in astronomical contexts was established by \citet{burn}, who introduced the concept of the Faraday dispersion function to describe the distribution of polarised radiation along the line of sight. When polarised electromagnetic waves propagate through a magnetised plasma, their plane of polarisation rotates by an amount proportional to the product of the electron density and the magnetic field component along the line of sight.

In astronomical contexts, Faraday rotation manifests as a phase shift in the complex polarisation components, producing a characteristic sinusoidal variation in the Stokes parameters $Q$ and $U$ as a function of wavelength squared ($\lambda^2$). The wavelength squared is a fundamental parameter in Faraday rotation studies, defined as $\lambda^2 = c^2/\nu^2$, where $c$ is the speed of light and $\nu$ is the frequency of the electromagnetic wave. This transformation from frequency to wavelength squared space is crucial because Faraday rotation effects manifest directly in $\lambda^2$-space, making it the natural coordinate system for analysing polarised emission.

This behaviour is elegantly captured in the mathematical formalism:

\begin{equation}
P(\lambda^2) = |P(\lambda^2)| e^{2i\chi(\lambda^2)} = Q(\lambda^2) + iU(\lambda^2),
\label{eq:p-pol}
\end{equation}

where the complex polarisation $P(\lambda^2)$ encodes both the magnitude and orientation of the polarised emission. The polarisation angle $\chi(\lambda^2)$ evolves linearly with wavelength squared:

\begin{equation}
\chi(\lambda^2) = \chi_0 + \text{RM} \lambda^2.
\end{equation}

The Rotation Measure (RM) serves as a key diagnostic tool, quantifying the cumulative Faraday rotation along the line of sight through the relationship:

\begin{equation}
\text{RM} = 0.81 \int_{\text{source}}^{\text{observer}} n_e(l)\, B_{\parallel}(l)\, dl,
\label{eq:rm}
\end{equation}

where RM has units of rad\,m$^{-2}$, $n_e(l)$ is the free-electron density (cm$^{-3}$), $B_{\parallel}(l)$ is the magnetic-field component parallel to the line of sight ($\mu$G), and $dl$ is the differential path length (pc). Because the integral is evaluated over the full path from source to observer, RM represents the total Faraday rotation accumulated along the entire line of sight and measured at the observer. By contrast, Faraday depth $\phi$ is defined at a specific position along that path: it is the same integral, but truncated at that position instead of extending all the way to the observer. Therefore, $\phi$ can vary with distance through the medium, while RM is the total value at the observer. In the special case where one region dominates the rotation, the measured RM is approximately equal to the Faraday depth of that region.


\subsection{Types of Faraday Rotation Sources}
\label{subsec:types-faraday}

In astrophysical contexts, we encounter various types of sources that exhibit different Faraday rotation characteristics. These can be broadly categorised into thin and thick sources, each providing unique insights into cosmic magnetic fields. The distinction between these source types is particularly important as they manifest differently in Faraday depth space and require different observational strategies.

\begin{figure}[ht!]
\centering
\includegraphics[width=\textwidth, keepaspectratio]{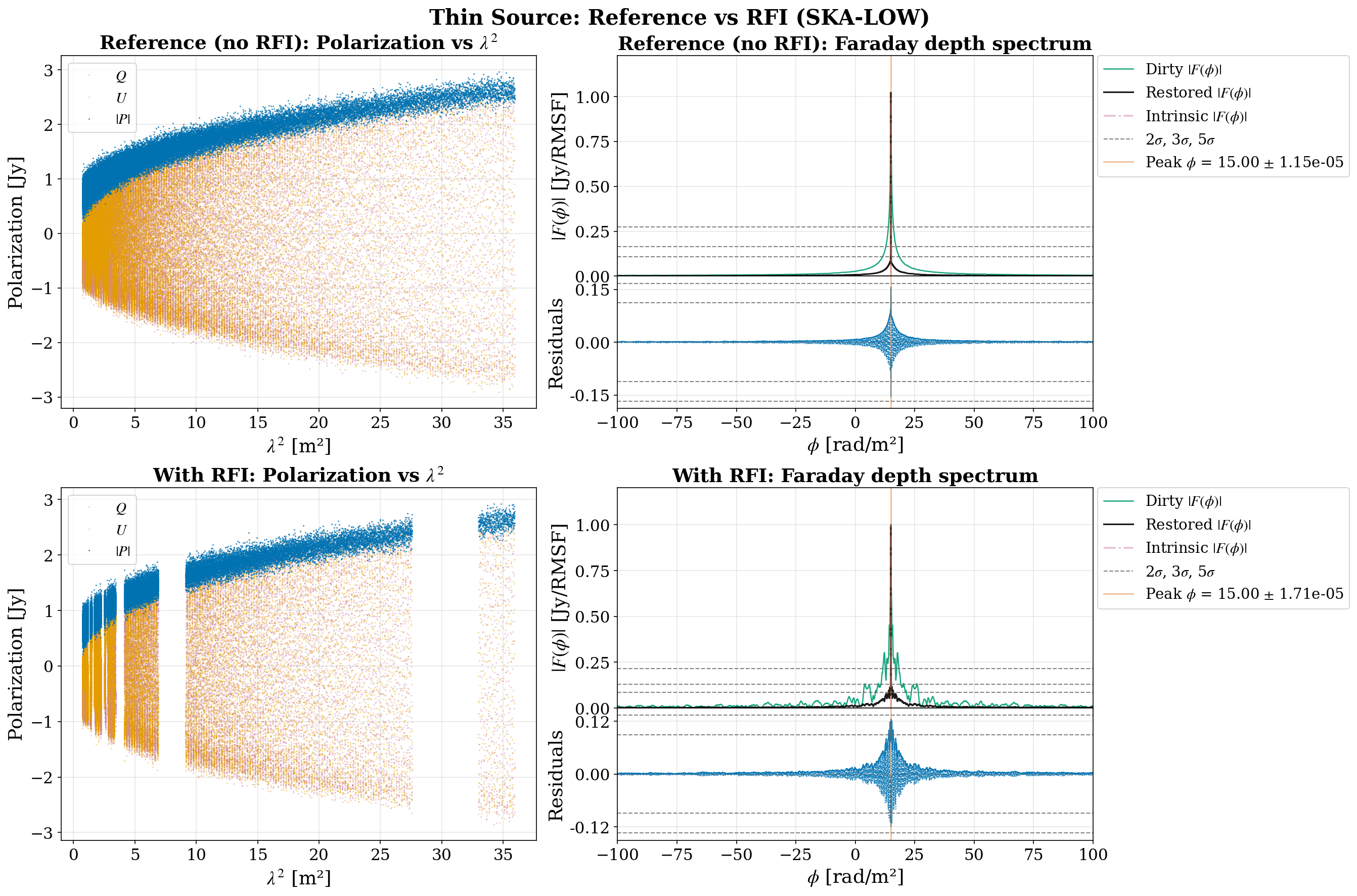}
\caption{Thin source observed with SKA-Low (50--350 MHz). \textbf{Top row:} reference (no RFI). \textbf{Bottom row:} with simulated RFI (30\% of channels removed in contiguous segments to mimic flagging). \textbf{Left:} polarisation (in Jy) vs $\lambda^2$. \textbf{Right:} Faraday depth spectrum including the intrinsic model (ground truth; pink dash-dotted curve), dirty and restored $|F(\phi)|$, and residuals (grey dashed lines indicate 2$\sigma$, 3$\sigma$, and 5$\sigma$ in Faraday depth). Input: Faraday depth $\phi = 15\,\mathrm{rad}\,\mathrm{m}^{-2}$, peak intensity 1.0 Jy. In both the reference and RFI cases the recovered peak is close to 1.0 Jy. The exacerbation of the sidelobes in the RFI case increases the noise in Faraday depth space and therefore the uncertainty on the peak; the restored spectrum shows stronger sidelobes and structured residuals, illustrating how gaps in $\lambda^2$ coverage degrade reconstruction quality.}
\label{fig:thin_source}
\end{figure}

Thin Faraday rotation sources are those where the emission and rotation occur in separate regions along the line of sight. A classic example is a distant radio galaxy or quasar, where the polarised synchrotron emission occurs in the source itself, and the Faraday rotation happens in the intervening medium (such as galaxy clusters, the Milky Way's interstellar medium, or the intergalactic medium). In Faraday depth space, these sources appear as sharp peaks, effectively delta functions, as the emission and rotation are well-separated (see Figure~\ref{fig:thin_source}). This makes them particularly useful for probing magnetic fields in the intervening medium, as the Faraday depth directly corresponds to the rotation measure of the medium.

\begin{figure}[ht!]
\centering
\includegraphics[width=\textwidth, keepaspectratio]{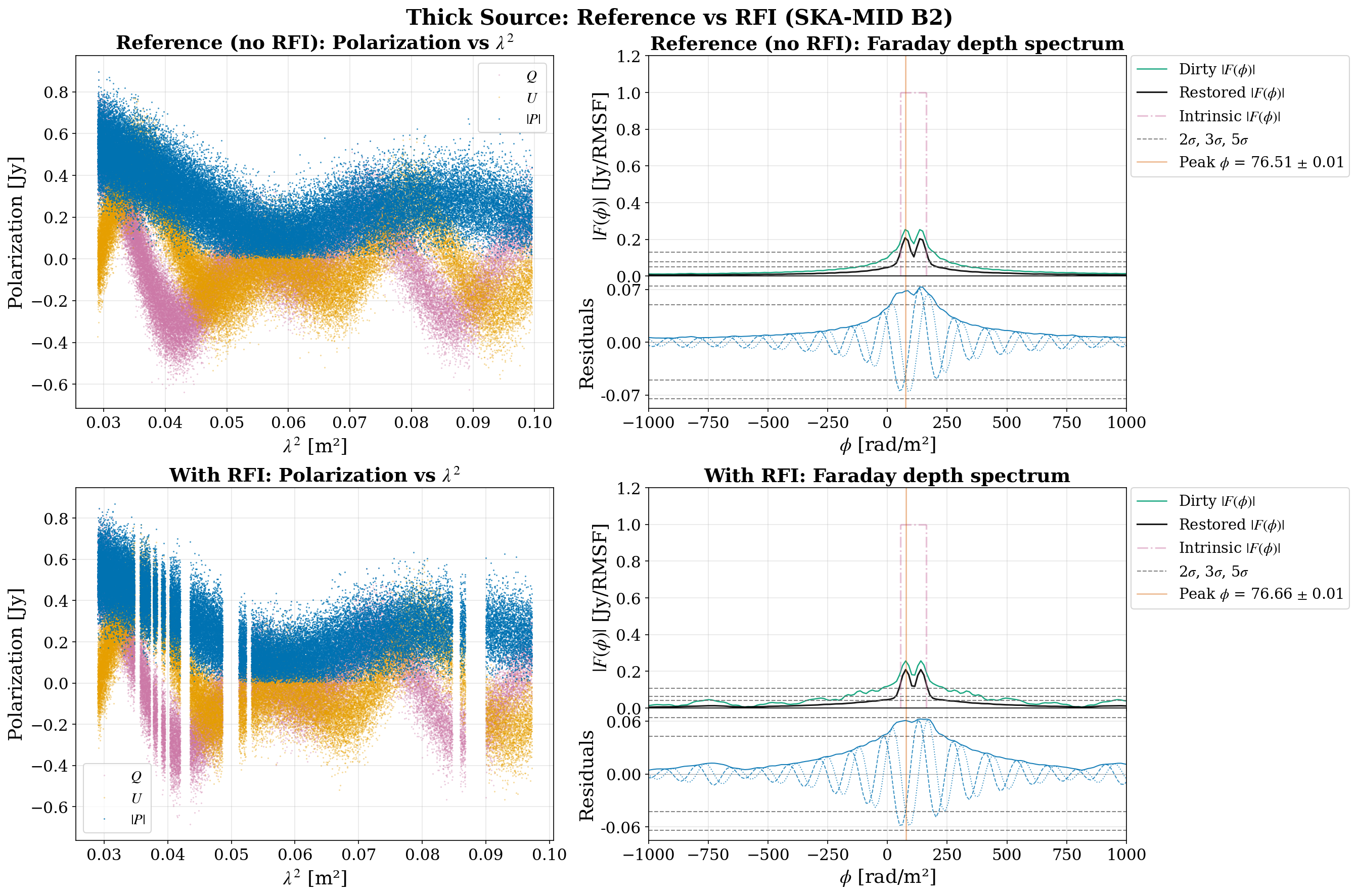}
\caption{Thick source observed with SKA-Mid Band 2 (950--1760 MHz). \textbf{Top row:} reference (no RFI). \textbf{Bottom row:} with simulated RFI (20\% of channels removed in contiguous segments to mimic flagging). \textbf{Left:} polarisation (in Jy) vs $\lambda^2$. \textbf{Right:} Faraday depth spectrum including the intrinsic model (ground truth; pink dash-dotted curve), dirty and restored $|F(\phi)|$, and residuals (grey dashed lines indicate 2$\sigma$, 3$\sigma$, and 5$\sigma$ in Faraday depth). Input: thick source (top-hat in Faraday depth, half-width $\approx 98\,\mathrm{rad}\,\mathrm{m}^{-2}$, centred at $\approx 196\,\mathrm{rad}\,\mathrm{m}^{-2}$), polarised intensity 1.0 Jy. For thick sources the emission is spread in Faraday depth space, so the peak in $|F(\phi)|$ does not reach 1.0 Jy. In both rows the broad distribution reflects the thick source structure; the exacerbation of the sidelobes in the RFI case increases the noise in Faraday depth space and the uncertainty on the recovered structure, illustrating how gaps in $\lambda^2$ coverage degrade reconstruction quality.}
\label{fig:thick_source}
\end{figure}

In contrast, thick Faraday rotation sources are those where emission and rotation occur in the same region. Examples include radio galaxy lobes, where relativistic electrons emit synchrotron radiation while mixed with thermal plasma; supernova remnants, which contain both relativistic electrons and magnetised thermal plasma; and galaxy clusters, where extended radio halos and relics exist within the magnetised intracluster medium. These sources appear as extended structures in Faraday depth space, often showing a sinc-like response in the Stokes Q and U parameters as a function of $\lambda^2$ (see Figure~\ref{fig:thick_source}). This extended structure provides information about both the field morphology and the emission mechanisms within the source itself.

\begin{figure}[ht!]
\centering
\includegraphics[width=\textwidth, keepaspectratio]{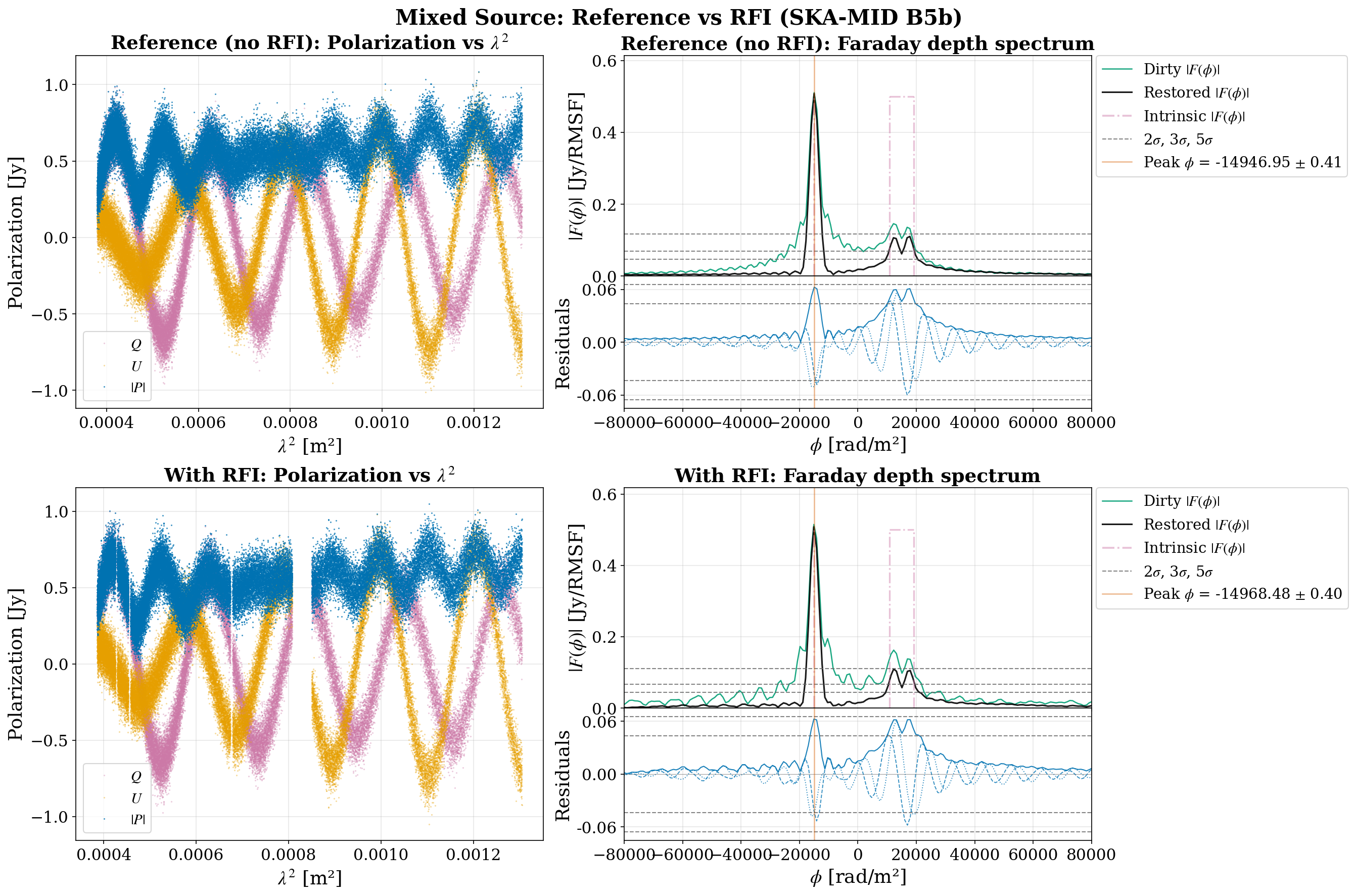}
\caption{Mixed source observed with SKA-Mid Band 5b (8.3--15.4 GHz). \textbf{Top row:} reference (no RFI). \textbf{Bottom row:} with simulated RFI (10\% of channels removed in contiguous segments to mimic flagging). \textbf{Left:} polarisation (in Jy) vs $\lambda^2$. \textbf{Right:} Faraday depth spectrum including the intrinsic model (ground truth; pink dash-dotted curve), dirty and restored $|F(\phi)|$, and residuals (grey dashed lines indicate 2$\sigma$, 3$\sigma$, and 5$\sigma$ in Faraday depth). Input: thin component (delta) at $\phi = -15.0\times10^{3}\,\mathrm{rad}\,\mathrm{m}^{-2}$, thick component (top-hat, half-width $4.1\times10^{3}\,\mathrm{rad}\,\mathrm{m}^{-2}$) centred at $\phi = +15.0\times10^{3}\,\mathrm{rad}\,\mathrm{m}^{-2}$; each component has peak polarised intensity 0.5 Jy. The thin (point) component peak reaches 0.5 Jy, but the thick component peak does not, confirming the same effect as in Figure~\ref{fig:thick_source}. The exacerbation of the sidelobes in the RFI case increases the noise and the uncertainty on the recovered structure. The complex morphology reflects the superposition of multiple magnetised regions with different properties along the line of sight.}
\label{fig:mixed_source}
\end{figure}

Mixed Faraday rotation sources represent a combination of both thin and thick components along the same line of sight. These sources can arise when multiple magnetised regions with different properties contribute to the observed polarisation, such as a background radio galaxy viewed through both a galaxy cluster (thick component) and the Milky Way's interstellar medium (thin component). In Faraday depth space, mixed sources exhibit complex structures that combine sharp peaks from thin components with extended distributions from thick components (see Figure~\ref{fig:mixed_source}), requiring careful analysis to disentangle the different contributions.

The distinction between thermal and non-thermal sources is also crucial in Faraday rotation studies. Non-thermal sources, such as radio galaxy jets, supernova remnants, and cosmic ray electrons in galaxy clusters, produce polarised synchrotron emission that can be modified by Faraday rotation. These sources are typically the ones we observe directly. On the other hand, thermal sources, such as HII regions and the hot gas in galaxy clusters, primarily act as Faraday rotating media. While they don't typically produce polarised emission themselves, they affect the polarisation of background sources, creating additional complexity in the Faraday depth structure.

\subsection{Observational Parameters}
The ability to resolve structures in Faraday depth space is governed by several fundamental parameters that depend on the $\lambda^2$ coverage of the observations. These parameters determine our capability to distinguish and characterise different magnetised structures along the line of sight:

\begin{itemize}
    \item The \textbf{Nominal Faraday Depth Resolution} ($\delta\phi_{\text{nom}}$) represents the minimum separation needed to distinguish two different Faraday components using the standard Brentjens \& de Bruyn (2005) approach:
    \begin{equation}
        \delta\phi_{\text{nom}} \approx \frac{2\sqrt{3}}{\lambda^2_{\text{max}} - \lambda^2_{\text{min}}},
    \end{equation}
    where $\lambda^2_{\text{max}}$ and $\lambda^2_{\text{min}}$ are the maximum and minimum wavelengths squared in the observation.

    \item The \textbf{Full Faraday Depth Resolution} ($\delta\phi_{\text{full}}$) represents the narrower resolution achievable by using the full complex spectrum without the $\lambda_0^2$ reference term \citep{Rudnick2023}. Here $\lambda_0^2$ is a reference wavelength squared often used when expressing the complex polarisation relative to a fixed wavelength; omitting it (i.e., not de-rotating to a reference $\lambda_0^2$) allows the full complex spectrum to contribute and yields the finer resolution given by
    \begin{equation}
        \delta\phi_{\text{full}} \approx \frac{2}{\lambda^2_{\text{max}} + \lambda^2_{\text{min}}}.
    \end{equation}
    This approach provides better discrimination of multiple Faraday components and reduces spurious features in Faraday tomography maps.

    \item The \textbf{Maximum Observable Faraday Depth} ($|\phi_{\text{max}}|$) defines the largest Faraday depth that can be measured without aliasing:
    \begin{equation}
        |\phi_{\text{max}}| \approx \frac{\sqrt{3}}{\delta\lambda^2},
    \end{equation}
    where $\delta\lambda^2$ is the width of a single frequency channel in $\lambda^2$ space.

    \item The \textbf{Largest Scale in Faraday Space} ($\phi_{\text{max-scale}}$) determines the maximum extent of resolved structures:
    \begin{equation}
        \phi_{\text{max-scale}} \approx \frac{\pi}{\lambda^2_{\text{min}}}.
    \end{equation}
\end{itemize}

These parameters are crucial for understanding the capabilities and limitations of Faraday rotation studies. The SKA's wide multi-frequency coverage (see Table~\ref{tab:aa-star-params}) will enable unprecedented resolution in Faraday depth space, allowing us to probe both thin and thick Faraday structures across various cosmic environments.

\subsection{Faraday Rotation in the SKA}

The phased deployment of the SKA will progressively enhance our capability to study cosmic magnetism through Faraday rotation measurements. The AA* configuration, operating with 144 dishes (80 SKA and 64 MeerKAT), will provide early science opportunities while the array is still expanding toward the full AA4 configuration with 197 dishes.

The different frequency bands available to the SKA enable multi-scale Faraday tomography, with each band offering different trade-offs between Faraday depth resolution and the maximum observable Faraday depth. These capabilities are particularly important given the diverse nature of Faraday rotation sources introduced in Section~\ref{subsec:types-faraday}. The ability to probe both thin sources (where emission and rotation are well-separated) and thick sources (where they occur in the same region) requires careful optimisation of observational parameters across different frequency ranges.

Table~\ref{tab:aa-star-params} summarises the key observational parameters for Faraday rotation studies using the AA* configuration across multiple frequency bands. The choice of frequency band depends on the type of source being studied: for thin sources probing the intervening medium, fine Faraday depth resolution is crucial, while for thick sources within magnetised plasmas, the maximum observable depth becomes more important.

\begin{table}[ht!]
\centering
\caption{Faraday rotation observational parameters for the AA* configuration.}
\label{tab:aa-star-params}
\begin{tabular}{lccccc}
\toprule
Band & Frequency & $\delta\phi_{\text{nom}}$ & $\delta\phi_{\text{full}}$ & $|\phi|_{\text{max}}$ & $\phi_{\text{max-scale}}$ \\
 & (MHz) & [rad m$^{-2}$] & [rad m$^{-2}$] & [rad m$^{-2}$] & [rad m$^{-2}$] \\
\midrule
SKA-Low      & 50--350 & $9.84\times10^{-2}$ & $5.45\times10^{-2}$ & $3.78\times10^{3}$ & $4.28\times10^{0}$ \\
SKA-Mid B1 & 350--1050 & $5.31\times10^{0}$ & $2.45\times10^{0}$ & $1.38\times10^{5}$ & $3.85\times10^{1}$ \\
SKA-Mid B2   & 950--1760 & $4.91\times10^{1}$ & $1.56\times10^{1}$ & $1.48\times10^{6}$ & $1.08\times10^{2}$ \\
SKA-Mid B5a  & 4600--8500 & $1.15\times10^{3}$ & $3.64\times10^{2}$ & $1.67\times10^{8}$ & $2.53\times10^{3}$ \\
SKA-Mid B5b  & 8300--15400 & $3.74\times10^{3}$ & $1.19\times10^{3}$ & $9.89\times10^{8}$ & $8.29\times10^{3}$ \\
\bottomrule
\end{tabular}
\end{table}

As shown in Table~\ref{tab:aa-star-params}, SKA-Low (50--350 MHz) offers the finest full Faraday depth resolution at 0.0545 rad m$^{-2}$ (with a nominal resolution of 0.0984 rad m$^{-2}$), making it ideal for detailed studies of magnetic field structures with very fine-scale features. This low-frequency band achieves a maximum observable Faraday depth of 3,783 rad m$^{-2}$, enabling the detection of sources with moderate to high rotation measures.

SKA-Mid Band 1 (350--1050 MHz) offers the finest Faraday depth resolution among the dish bands (full resolution $\sim$2.5 rad m$^{-2}$) with a high maximum observable depth ($\sim$138,000 rad m$^{-2}$), bridging SKA-Low and the higher-frequency Mid bands. Band 2 offers moderate resolution (15.6 rad m$^{-2}$ full resolution) with a high maximum observable depth, suitable for studying extended magnetic field structures. Band 5a has coarser resolution (364 rad m$^{-2}$) but extraordinarily high maximum depth (167 million rad m$^{-2}$), while Band 5b trades Faraday depth resolution (1188 rad m$^{-2}$, coarsest of the Mid bands) for higher angular resolution on the sky and sensitivity at the highest frequencies. This multi-band approach allows astronomers to select the optimal frequency range for their specific scientific objectives, whether investigating detailed magnetic field morphology on small scales or probing the large-scale magnetic structure of the Universe.

The progression from AA* to full AA4 will further enhance these capabilities, with improved sensitivity enabling the detection of fainter polarised sources and higher-quality measurements of rotation measure structures across all frequency bands.

\subsubsection{Depolarisation}

In Faraday rotation studies, depolarisation refers to the decrease of observed polarised intensity with wavelength squared. It can arise in several ways: within a channel, different frequencies experience different Faraday rotation and the polarised signal is effectively averaged (bandwidth depolarisation); along the line of sight, multiple Faraday components can interfere (depth depolarisation); or the magnetic field may vary across the beam and cause cancellation (beam depolarisation).

For bandwidth depolarisation, the observed fractional polarisation $p(\lambda^2)$ is given by
\begin{equation}
p(\lambda^2) = p_0 \, \mathrm{sinc}(\phi \, \Delta\lambda^2) = p_0 \frac{\sin(\phi \, \Delta\lambda^2)}{\phi \, \Delta\lambda^2},
\label{eq:bandwidth_depolarisation}
\end{equation}
where $p_0$ is the intrinsic fractional polarisation, $\phi$ is the Faraday depth (rad\,m$^{-2}$) characterising the rotation across the channel, and $\Delta\lambda^2$ is the channel width in $\lambda^2$ space. For depth depolarisation, where there is a distribution of Faraday depths along the line of sight, the depolarisation follows
\begin{equation}
p(\lambda^2) = p_0 e^{-2\sigma_{\phi,\mathrm{dep}}^2\lambda^4},
\label{eq:depth_depolarisation}
\end{equation}
where $\sigma_{\phi,\mathrm{dep}}$ is the standard deviation of the Faraday depth distribution used in the depolarisation model. This depth depolarisation relation applies to a non-synchrotron-emitting medium lying between the observer and the polarised source (e.g.\ foreground thermal plasma).

\begin{figure}[ht!]
\centering
\includegraphics[width=\textwidth, keepaspectratio]{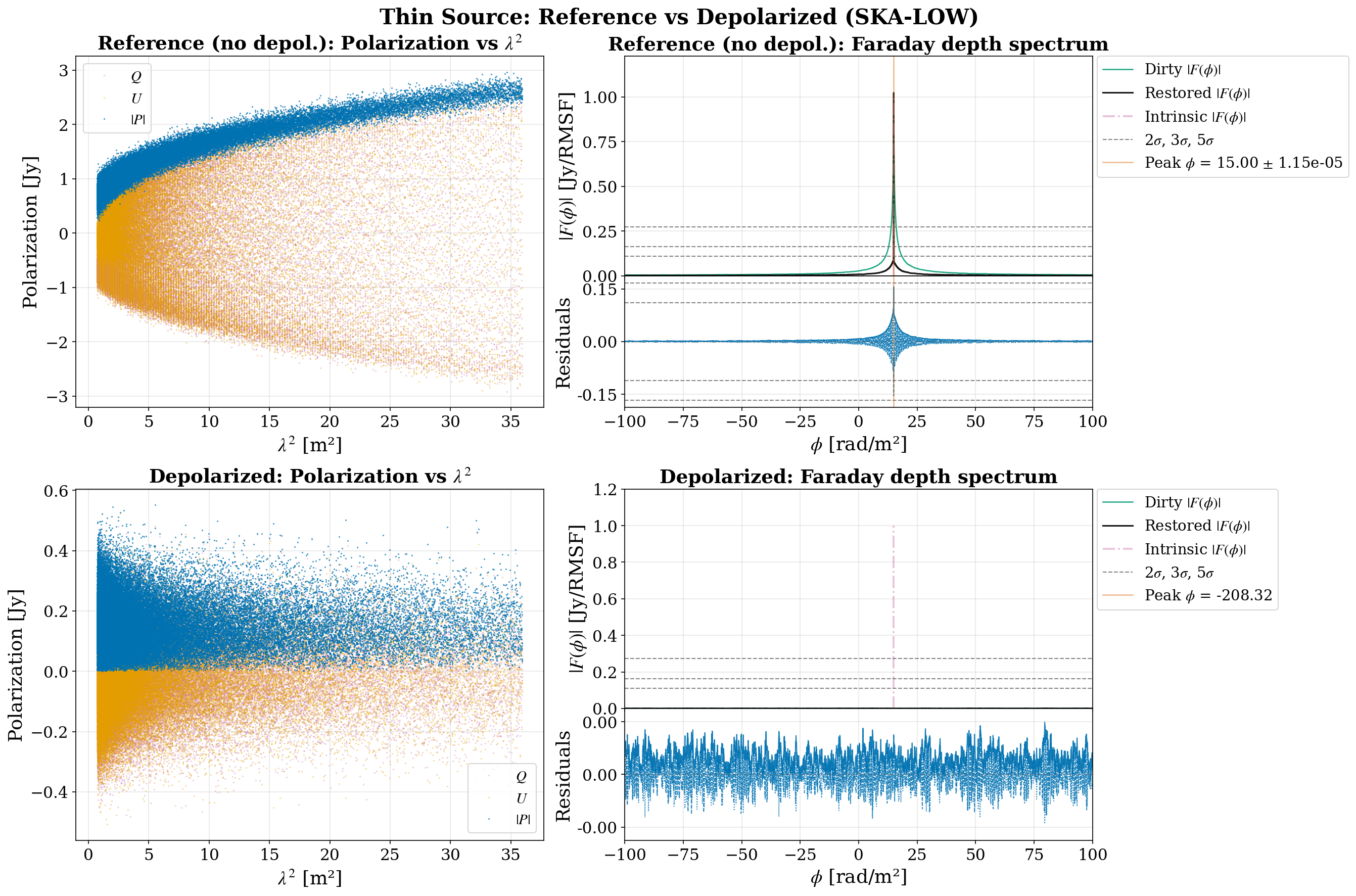}
\caption{Depolarisation for a thin source with SKA-Low (50--350 MHz). \textbf{Top row:} reference (no depolarisation). \textbf{Bottom row:} with depolarisation. \textbf{Left:} polarisation (in Jy) vs $\lambda^2$. \textbf{Right:} Faraday depth spectrum including the intrinsic model (ground truth; pink dash-dotted curve), dirty and restored $|F(\phi)|$, and residuals (grey dashed lines indicate 2$\sigma$, 3$\sigma$, and 5$\sigma$ in Faraday depth). Input: $\phi = 15\,\mathrm{rad}\,\mathrm{m}^{-2}$, peak intensity 1.0 Jy. In the reference case the peak is recovered with high precision. With depolarisation the polarised signal is strongly attenuated, the Faraday depth spectrum becomes noise-dominated, the recovered peak is incorrect and the uncertainty increases sharply, illustrating that at low frequencies depolarisation can severely compromise both the polarised signal and the accuracy and precision of RM recovery.}
\label{fig:depolarisation_low}
\end{figure}

\begin{figure}[ht!]
\centering
\includegraphics[width=\textwidth, keepaspectratio]{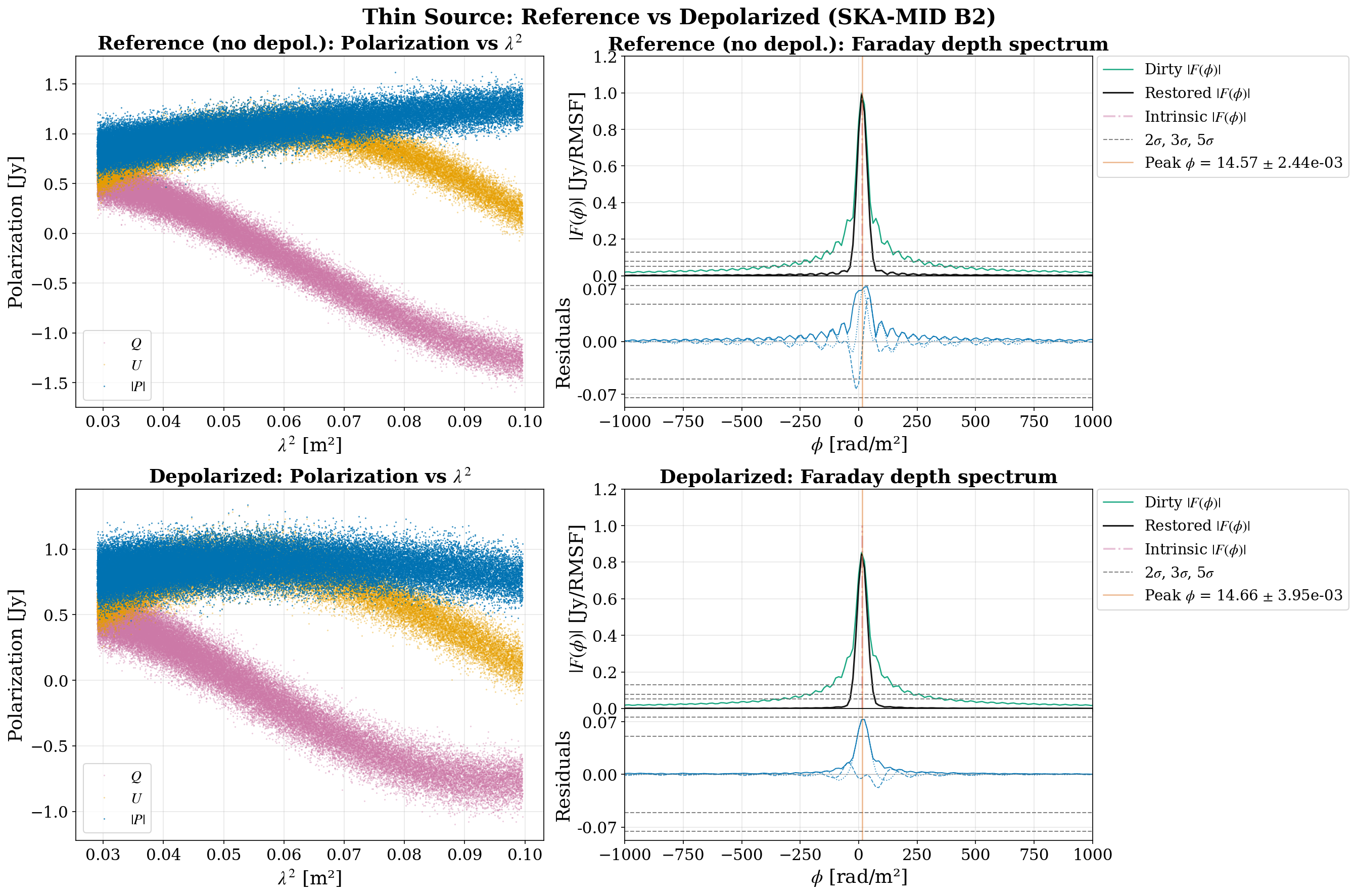}
\caption{Depolarisation for a thin source with SKA-Mid Band 2 (950--1760 MHz). \textbf{Top row:} reference (no depolarisation). \textbf{Bottom row:} with depolarisation. \textbf{Left:} polarisation (in Jy) vs $\lambda^2$. \textbf{Right:} Faraday depth spectrum including the intrinsic model (ground truth; pink dash-dotted curve), dirty and restored $|F(\phi)|$, and residuals (grey dashed lines indicate 2$\sigma$, 3$\sigma$, and 5$\sigma$ in Faraday depth). Input: $\phi = 15\,\mathrm{rad}\,\mathrm{m}^{-2}$, peak intensity 1.0 Jy (same as Figure~\ref{fig:depolarisation_low}). Depolarisation is less severe at this band; however, the amplitude of the polarised signal is attenuated in the dirty spectrum and the reconstruction. Compared to SKA-Low (Figure~\ref{fig:depolarisation_low}), depolarisation is less severe here, but the resolution-amplitude trade-off remains.}
\label{fig:depolarisation_b5b}
\end{figure}

The frequency dependence of depolarisation is particularly important. Because Faraday rotation scales with $\lambda^2$, lower frequencies are more susceptible to depolarisation. At SKA-Low, depolarisation can strongly attenuate the polarised signal and dominate the Faraday depth spectrum, so that both the polarised amplitude and the accuracy and precision of RM recovery are compromised (Figure~\ref{fig:depolarisation_low}). At higher frequencies (e.g.\ SKA-Mid Band 2), depolarisation is less severe, but the coarser Faraday depth resolution makes it harder for reconstruction algorithms to recover the polarised amplitude perfectly; depolarisation worsens this effect, as seen in the reduced signal even in the dirty spectrum (Figure~\ref{fig:depolarisation_b5b}). Figures~\ref{fig:depolarisation_low} and~\ref{fig:depolarisation_b5b} illustrate this frequency-dependent trade-off between depolarisation at low frequencies and resolution at higher frequencies.

\subsection{Faraday Measurement Synthesis}

We use the term Faraday Measurement Synthesis to mean the family of techniques that reconstruct the Faraday dispersion function $F(\phi)$ from observed complex polarised intensity $P(\lambda^2)$. This includes rotation measure synthesis (RMS), RM-CLEAN, QU-fitting, and related inversion methods; the terms ``RM synthesis'' and ``Faraday tomography'' are more common in the literature and are largely interchangeable.

The inverse problem is difficult for several reasons. Interferometers record data at linearly spaced frequencies, so sampling in $\lambda^2$ is irregular and non-uniform. Flagging of channels affected by radio frequency interference (RFI) introduces further gaps (see Figures~\ref{fig:thin_source}, \ref{fig:thick_source}, and~\ref{fig:mixed_source}). These gaps reduce the effective $\lambda^2$ coverage, broaden the rotation measure spread function (RMSF), and increase the uncertainty on the recovered Faraday depth; for a simple thin source (a delta-function in Faraday depth), the inferred RM can be biased or less precise even when RMS methods are used. In addition, we only observe $P(\lambda^2)$ for $\lambda^2 > 0$. As a result, the inversion from $P(\lambda^2)$ to $F(\phi)$ is ill-posed: many different Faraday depth structures can produce similar observed $P(\lambda^2)$, so the solution is not unique and small changes in the data can cause large changes in the reconstructed $F(\phi)$. Reliable reconstructions therefore require sophisticated mathematical and computational techniques.

To illustrate these challenges, the simulated data and the corresponding reconstructions in Figures~\ref{fig:thin_source}, \ref{fig:thick_source}, \ref{fig:mixed_source}, \ref{fig:depolarisation_low}, and~\ref{fig:depolarisation_b5b} were generated with \texttt{cs-romer} and with our RM-CLEAN implementation, respectively, both within \texttt{cs-romer} \citep{csromer}. We then added synthetic noise, requiring a minimum polarised signal-to-noise ratio (SNR) of 10 in the worst case, and applied band-dependent scaling factors (1.2 for SKA-Low, 1.0 for Band 2, and 0.9 for Band 5b) to the injected noise amplitude to emulate band-specific sensitivity and channelisation conditions in the simulated data. In addition, we simulated RFI by flagging channels in contiguous segments (30\% for SKA-Low, 20\% for Band 2, and 10\% for Band 5b). In all Faraday-depth plots we overplot the intrinsic model (ground truth) as a pink dash-dotted curve, together with the dirty and restored reconstructions. For the thin-source depolarisation examples shown in Figures~\ref{fig:depolarisation_low} and~\ref{fig:depolarisation_b5b}, we used depth depolarisation with $\sigma_{\phi,\mathrm{dep}} = 5\,\mathrm{rad}\,\mathrm{m}^{-2}$. Throughout this chapter, ``thin'' and ``thick'' are defined by Faraday-depth extent relative to the RMSF (thin: unresolved/delta-like; thick: resolved/extended). The RM error, denoted $\sigma_{\mathrm{RM}}$ in the figure legends, is computed from $\sigma_{\mathrm{RM}} \approx \mathrm{FWHM}_{\mathrm{RMSF}} / (2\,\mathrm{SNR})$, where $\mathrm{FWHM}_{\mathrm{RMSF}}$ is the full width at half maximum of the RMSF and SNR is the signal-to-noise ratio of the peak. Finally, the peak signal-to-noise ratio (PSNR) is defined as $\mathrm{PSNR} = |P_{\mathrm{peak}}| / \sigma_{\mathrm{noise}}$, where $P_{\mathrm{peak}}$ is the peak amplitude corrected for Ricean bias and $\sigma_{\mathrm{noise}}$ is estimated from the residuals. For a comprehensive review of Faraday tomography and its applications, see \citet{cosmic_magnetism_review}.

Mathematically, the relationship between the observed complex polarisation and the Faraday dispersion function is given by a Fourier-like transform:
\begin{equation}
    P(\lambda^2) = \int_{-\infty}^{+\infty} F(\phi) \, e^{2i\phi\lambda^2} \, d\phi,
    \label{eq:forward}
\end{equation}
where \( P(\lambda^2) \) is the observed complex polarisation as a function of wavelength squared, and \( F(\phi) \) is the Faraday dispersion function, representing the distribution of polarised emission as a function of Faraday depth \( \phi \).

The inverse problem (recovering \( F(\phi) \) from the observed \( P(\lambda^2) \)) is formally given by:
\begin{equation}
    F(\phi) = \frac{1}{\pi} \int_{0}^{+\infty} P(\lambda^2) \, e^{-2i\phi\lambda^2} \, d\lambda^2,
    \label{eq:inverse}
\end{equation}
where the lower limit of the integral reflects the fact that we only have measurements for \( \lambda^2 > 0 \).

In practice, the limited and irregular sampling in \( \lambda^2 \) space, as well as the presence of noise and RFI, means that the inversion is not straightforward. Thus, the resulting \( F(\phi) \) is convolved with a window function (the Rotation Measure Spread Function, RMSF), analogous to the point spread function in imaging, which limits the resolution and fidelity of the Faraday depth reconstruction. Advanced deconvolution and regularisation techniques are therefore required to recover an estimate of the true Faraday structure from the observed data.

\subsection{Faraday Measurement Reconstruction Methods}

Until now, we have formulated the problem as a 1D problem over a single line of sight. Several methods have been developed and implemented to solve this inverse problem, ranging from simple analytical approaches to sophisticated numerical techniques. However, we will see that this formulation is not entirely correct, as we should use the $(u,v,w)$ components as well in our transforms to properly account for the three-dimensional nature of the problem. Here we split our discussion into two main approaches: 1D methods and 3D methods.

\subsubsection{1D Methods}

The 1D formulation treats Faraday Measurement Synthesis as a one-dimensional inverse problem along the line of sight. This approach assumes that the Faraday depth structure can be adequately described by considering only the $\lambda^2$ dependence of the complex polarisation, effectively treating each line of sight independently. While this simplification has been widely used and has provided valuable insights into cosmic magnetism, it has inherent limitations that become apparent when dealing with extended sources or complex magnetic field geometries.

Several techniques have been developed within this 1D framework, each addressing different aspects of the inverse problem and offering various trade-offs between computational efficiency, accuracy, and robustness to noise.

\textbf{Rotation Measure Synthesis (RMS)} represents the most fundamental approach to Faraday Measurement Synthesis, analogous to obtaining the ``dirty image'' in radio interferometric imaging. This method directly applies the inverse Fourier transform to recover the Faraday dispersion function from the observed complex polarisation, without any deconvolution or cleaning steps. The technique was first introduced by \citet{brentjens} and has since become the standard tool for Faraday rotation studies. RMS provides a direct, non-parametric reconstruction of the Faraday depth structure, making it particularly valuable for initial exploration of unknown or complex magnetic field geometries. However, just like the dirty image in radio imaging, the method suffers from limited resolution due to the finite $\lambda^2$ coverage and the presence of the Rotation Measure Spread Function (RMSF), which acts as a convolution kernel that blurs the true Faraday structure. The RMSF is the Faraday equivalent of the point spread function (PSF) in imaging, and the resulting ``dirty'' Faraday depth spectrum contains sidelobes and artefacts that limit the fidelity of the reconstruction. In practice, RMS and CLEAN-based deconvolution are often used together: RMS yields the dirty Faraday depth spectrum, and a Hogbom-like 1D CLEAN is then applied to deconvolve the RMSF and recover discrete components.

\textbf{QU-fitting} represents a parametric approach that fits specific physical models to the observed Stokes Q and U parameters as a function of $\lambda^2$. This method, pioneered by \citet{qu-fitting-1}, assumes that the Faraday depth structure can be described by a limited number of discrete components, each characterised by specific parameters such as Faraday depth, intrinsic polarisation angle, and fractional polarisation. QU-fitting has been further developed and applied in various contexts \citep{qu-fitting-2, qu-fitting-3, qu-fitting-4, 10.1093/mnras/sty2862}. QU-fitting offers several advantages over RMS, including better noise handling, the ability to separate multiple Faraday components, and the provision of uncertainty estimates for the fitted parameters. However, the method requires prior knowledge or assumptions about the number and nature of Faraday components, which may not always be available or accurate.

\textbf{CLEAN-based algorithms} attempt to remove the effects of the RMSF to recover the true Faraday depth structure. These methods, inspired by the CLEAN algorithm used in radio imaging, employ a greedy/brute force approach to iteratively identify and remove sidelobes from the dirty Faraday spectrum (typically obtained from RMS in a preceding step). The RM-CLEAN technique has been extensively studied and applied \citep{heald, 10.1093/pasj/psu030, 10.1093/pasj/psw039}. These methods work best for thin Faraday sources and can significantly improve the resolution and fidelity of Faraday depth reconstructions for such sources. However, they suffer from several important limitations. First, they can be computationally intensive, particularly for large datasets. Second, they are highly sensitive to noise and the quality of the $\lambda^2$ coverage. Third, they may produce false signals when sources are closely separated in Faraday depth space, a problem that becomes more severe as the separation approaches the resolution limit. Fourth, they can be affected by the intrinsic polarisation angle ambiguity: at a given Faraday depth, complex polarisation from different physical depths along the line of sight is averaged, so that different combinations of intrinsic polarisation angles can produce similar Faraday depth structures, making the solution degenerate. Finally, they may fail to converge properly when dealing with extended or complex Faraday structures, as the algorithm assumes point-like sources.

\textbf{Regularisation methods} have been developed to address the ill-posed nature of the Faraday Measurement Synthesis inverse problem by incorporating prior information about the expected characteristics of the Faraday depth structure. These techniques employ mathematical constraints such as sparsity (assuming few non-zero components) or smoothness (assuming gradual variations) to stabilise the inversion process and produce more physically meaningful reconstructions. Unlike CLEAN-based methods that assume point-like sources, regularisation approaches can effectively reconstruct thin, thick, and mixed Faraday sources by leveraging these prior characteristics. Methods such as sparse reconstruction techniques \citep{Li, Li2, Andrecut_2011,akiyama2018faraday,csromer} have been successfully applied to Faraday rotation studies. However, while regularisation methods can effectively suppress noise and reduce artefacts in the reconstructed Faraday depth structure, they require careful tuning of regularisation parameters and may introduce biases depending on the chosen prior assumptions.

These 1D methods have been successfully applied to numerous astronomical sources and have provided important constraints on magnetic field properties in various cosmic environments. However, they suffer from fundamental limitations when dealing with extended sources or when the magnetic field structure has significant transverse variations. In practice, these 1D methods are typically applied using a two-step approach: first reconstructing Stokes Q and U cubes as a function of frequency through traditional aperture synthesis imaging at the same resolution, and then applying the 1D Faraday Measurement Synthesis techniques to each line of sight independently. This 2+1D approach, while widely used, introduces several significant limitations including the necessity to match resolutions between different frequency images, the compounding of artefacts through multiple deconvolution steps, and the inefficient use of available data that reduces overall sensitivity and degrades image fidelity.

\subsubsection{3D Methods}

The 3D formulation, pioneered by \citet{bell_synergy}, addresses these limitations by directly imaging the Faraday spectrum from interferometric visibility data. This approach, termed ``Faraday synthesis'', combines aperture synthesis and rotation measure synthesis into a single algorithm, avoiding the traditional two-step process entirely.

Faraday synthesis directly relates the Faraday spectrum to the visibilities of linearly polarised emission through a 3D Fourier transformation that simultaneously accounts for the $(u,v,\lambda^2)$ sampling of the interferometric data. This results in a 3D dirty beam that properly describes the instrument response and provides several key advantages:

\begin{itemize}
    \item \textbf{Improved resolution}: Achieves better angular resolution by utilizing the full $(u,v)$ coverage without requiring data downweighting to match resolutions across frequencies.
    \item \textbf{Enhanced sensitivity}: Uses all available data across the full bandwidth during imaging and deconvolution, resulting in approximately 20\% lower noise levels compared to traditional methods.
    \item \textbf{Reduced artefacts}: Eliminates the compounding of errors from multiple deconvolution steps, significantly reducing spurious sources and improving dynamic range.
    \item \textbf{Better foundation for advanced algorithms}: Provides a more accurate description of the instrument response, enabling the development of sophisticated reconstruction techniques based on proper noise modeling.
\end{itemize}

The approach requires specialised 3D CLEAN algorithms (such as FSCLEAN and the more recent DDFSCLEAN) that operate directly in the sky-Faraday depth space $(l, m, \phi)$ rather than working sequentially in spatial and frequency dimensions. These algorithms naturally account for bandwidth depolarisation effects, allowing accurate recovery of polarised flux across a wide range of Faraday depths \citep{faraday_synthesis_direction_dependent}.

Modern implementations have extended the 3D approach to incorporate direction-dependent effects through integration with faceting frameworks \citep{faraday_synthesis_direction_dependent}, enabling accurate polarised imaging in the presence of instrumental and ionospheric effects. When applied to both simulated and observational data, the 3D approach consistently demonstrates deeper deconvolution, reduced artefacts around bright sources, improved dynamic range, and better recovery of source parameters, enabling detection of polarised sources missed by traditional surveys.

The 1D and 3D methods described in this section are the basis for extracting RMs in the systematic surveys (RM grids) discussed in the next section. The 3D approach in particular requires more sophisticated computational techniques and larger memory for the full Faraday cube, but as computing resources and algorithms advance, it holds significant promise for studies with the SKA. When applied to SKA data, these methods benefit from wide fractional bandwidth and multi-band coverage (SKA-Low to Band 5b), which provide excellent $\lambda^2$ coverage and Faraday depth resolution, and from high sensitivity, which allows fainter polarised sources to be used for RM grids and improves the precision of recovered Faraday depths. Real observations will, however, face three practical constraints. First, radio frequency interference: despite remote sites, low frequencies remain more susceptible to RFI, so that flagging introduces gaps in $\lambda^2$, broadens the RMSF, and increases the uncertainty on the recovered RM (Figures~\ref{fig:thin_source}--\ref{fig:mixed_source}). Second, at the lowest frequencies (SKA-Low), depolarisation can strongly attenuate the polarised signal and limit the accuracy of RM recovery (Figures~\ref{fig:depolarisation_low} and~\ref{fig:depolarisation_b5b}). Third, data volume: in 1D RM synthesis one works with Stokes Q and U cubes $(l, m, \lambda^2)$, and in 3D with $(u, v, \lambda^2)$; the fine spectral and angular resolution of the SKA therefore implies very large data volumes (often many terabytes per field), so that big data management libraries and tools are a necessity. These methods and practical considerations will shape how RM grids are constructed and analysed with the SKA, as we discuss in the next section.

\section{RM Grids in the SKA Era}

Rotation Measure (RM) grids are a key observational tool for studying cosmic magnetism on large scales, and one of the most important methods for probing magnetic fields in galaxy clusters, the Milky Way's interstellar medium, and the intergalactic medium. This section outlines what an RM grid is, how such grids are constructed, their main applications, and the prospects with the SKA.

\subsection{What is an RM Grid?}

An RM grid is a systematic survey of rotation measures over a region of sky. Unlike a single RM, which probes one line of sight, a grid samples many lines of sight and thus provides spatial information on the intervening magnetic field. Each background polarised source yields one measurement of the integrated line-of-sight magnetic field through Equation~\ref{eq:rm}; combining many such measurements produces a map that encodes the field's strength, orientation, and coherence scale. RM grids are typically built toward extended targets such as galaxy clusters, supernova remnants, or large-scale Galactic structure, where this spatial sampling is most informative.

\subsection{Construction of RM Grids}

The construction of an RM grid involves several key steps, each requiring careful observational and analysis techniques. The process begins with the identification and selection of suitable background polarised sources. These sources must be sufficiently bright in polarised emission to enable reliable RM measurements, and they should be distributed across the region of interest to provide adequate spatial sampling. The density of sources in the grid determines the spatial resolution of the resulting magnetic field map, with higher source densities enabling finer resolution.

Once suitable sources are identified, observations are conducted across a wide frequency range to enable accurate RM determination. The SKA's multi-frequency capabilities, spanning from 350 MHz to 15.4 GHz, provide excellent coverage for RM grid construction. The wide bandwidth enables precise measurement of the wavelength-squared dependence of polarisation angle rotation, which is essential for accurate RM determination. For each source, the complex polarisation $P(\lambda^2)$ is measured across the available frequency range, and the RM is determined through either Rotation Measure Synthesis or parametric fitting techniques such as QU-fitting (as described in the section on Faraday Measurement Reconstruction Methods).

The quality of an RM grid depends critically on several factors. First, the number density of sources determines the spatial resolution of the grid. Higher source densities enable finer sampling of magnetic field structures, but require more extensive observations. Second, the accuracy of individual RM measurements depends on the signal-to-noise ratio of the polarised emission and the quality of the $\lambda^2$ coverage. The SKA's high sensitivity will enable RM measurements for fainter sources, increasing the achievable source density in RM grids. Third, the frequency coverage determines both the accuracy of RM measurements and the ability to detect and correct for depolarisation effects that can complicate RM determination.

For sources with complex Faraday depth structures (thick or mixed sources), the construction of RM grids becomes more challenging. In such cases, the simple linear relationship between polarisation angle and $\lambda^2$ breaks down, and more sophisticated analysis techniques are required. The SKA's wide frequency coverage and high sensitivity will enable the application of Faraday Measurement Synthesis techniques to extract reliable RM values even from sources with complex Faraday depth structures.

\subsection{Applications and Uses of RM Grids}

RM grids have become an essential tool for studying magnetic fields across a wide range of astrophysical environments. One of the most important applications is the study of magnetic fields in galaxy clusters. Cluster magnetic fields play crucial roles in various physical processes, including the confinement of cosmic rays, the suppression of heat conduction, and the evolution of cluster structure. RM grids constructed using background radio galaxies viewed through galaxy clusters reveal the magnetic field structure of the intracluster medium, providing constraints on field strength, coherence scale, and morphology \citep[e.g.][]{Anderson2021, Anderson2024, Loi01.2026.SKA}. Early POSSUM results have already demonstrated the power of denser RM grids: \citet{Anderson2021} used $\sim$25 RMs deg$^{-2}$ toward the Fornax cluster to reveal a large reservoir of magnetised plasma extending beyond the X-ray emission, and \citet{Anderson2024} used 22,817 RMs over 1520\,deg$^2$ to probe magnetised gas in 55 galaxy groups and the warm-hot intergalactic medium.

The analysis of RM grids in galaxy clusters typically involves statistical techniques to extract information about magnetic field properties. The dispersion of RMs across the grid provides constraints on the magnetic field strength, while the spatial correlation of RMs reveals information about the coherence scale of the magnetic field. The power spectrum of RM fluctuations can be used to infer the magnetic field power spectrum, providing insights into the origin and evolution of cluster magnetic fields, including dynamo processes in the intracluster medium \citep{AritraBasu01.2026.SKA}. The SKA will improve source density and spatial resolution for such studies.

RM grids are also essential for studying the magnetic field structure of the Milky Way's interstellar medium \citep{Taylor-2009, VanEck2011, faradaysky2020, Sun01.2026.SKA}. By measuring RMs for extragalactic sources distributed across the sky, we can construct all-sky RM grids that reveal the large-scale field morphology of our Galaxy. The NVSS RM catalogue provided the first $\sim$1 RM deg$^{-2}$ coverage over most of the northern sky \citep{Taylor-2009}; a consolidated catalogue of 55,819 RMs from 42 published catalogues \citep{VanEck2023} and subsequent full-sky RM maps \citep{faradaysky2020} have extended and refined Galactic magnetic field studies. These grids have revealed the spiral structure of the Galactic magnetic field, the presence of magnetic field reversals, and the complex three-dimensional structure of the Galactic magnetic field. The SKA will provide denser all-sky RM grids and better coverage of the Galactic plane, where source confusion and absorption currently limit observations.

Another important application of RM grids is the study of magnetic fields in supernova remnants and other extended Galactic sources. RM grids constructed using background sources viewed through supernova remnants reveal the field morphology of these objects, providing insights into the physics of shock acceleration and magnetic field amplification. The SKA's high angular resolution will support detailed RM grids for individual supernova remnants, revealing fine-scale field structure that is currently unresolved.

RM grids also play a crucial role in the study of magnetic fields in the intergalactic medium and cosmic web \citep{Carretti2022, Anderson2024}. By measuring RMs for sources at various redshifts, we can construct three-dimensional RM grids that probe magnetic fields across cosmic time. POSSUM and LOFAR are already being used to detect magnetised filaments and group haloes \citep{10.1093/mnras/stac3820,Anderson2024}. These measurements provide constraints on the origin and evolution of cosmic magnetic fields, addressing fundamental questions about when and how magnetic fields were first generated in the Universe.

\subsection{RM Grids with the SKA}

The applications described above set a baseline that the SKA will advance in sensitivity, angular resolution, and frequency coverage. Current sensitivity benchmarks are set by POSSUM ($\sim$18--25\,$\mu$Jy\,beam$^{-1}$ in Stokes Q and U \citep{Vanderwoude2024, GaenslerPOSSUM}) and by the LOFAR LoTSS RM grid (median detection threshold $\sim$0.6\,mJy\,beam$^{-1}$ at 120--168\,MHz \citep{10.1093/mnras/stac3820}). To quantify the improvement, we use the SKA Observatory sensitivity calculator\footnote{SKAO Sensitivity Calculator: Mid \texttt{https://sensitivity-calculator.skao.int/mid}, Low \texttt{https://sensitivity-calculator.skao.int/low}.}. For a representative RM-grid setup (16\,h integration, Band~2, 1.35\,GHz central frequency, 0.5\,GHz bandwidth, position 03:38, $-35^\circ27'$), SKA-Mid reaches $\sim$5.4\,$\mu$Jy\,beam$^{-1}$ (1\,$\sigma$) in the AA* configuration and $\sim$2.9\,$\mu$Jy\,beam$^{-1}$ at full AA4 (roughly 4 and 7 times fainter than current POSSUM depths, respectively), so that more background sources per square degree become usable for RM measurements. In the low-frequency regime, the SKA-Low calculator gives $\sim$27\,$\mu$Jy\,beam$^{-1}$ (AA*) and $\sim$15\,$\mu$Jy\,beam$^{-1}$ (AA4) for a comparable setup (16\,h, 145--200\,MHz, 150--300\,MHz bandwidth, position 06:00, $-40^\circ$ to avoid bright sources), or about 22 and 39 times fainter than the LoTSS median threshold---extending these gains to the finest Faraday depth resolution offered by SKA-Low and enabling RM-grid programmes aimed at the origin of cosmic magnetism \citep{OSullivan01.2026.SKA}.

In RM grid density, present benchmarks range from the NVSS RM catalogue at $\sim$1\,RM\,deg$^{-2}$ \citep{Taylor-2009} and LOFAR LoTSS at $\sim$0.43\,RMs\,deg$^{-2}$ \citep{10.1093/mnras/stac3820} to POSSUM at 30--50\,RMs\,deg$^{-2}$ over 20,630\,deg$^2$ with median RM uncertainties of $\sim$1\,rad\,m$^{-2}$ \citep{Vanderwoude2024, GaenslerPOSSUM}. Requirements depend on the target scale: cluster and group studies benefit from tens of RMs per square degree or more to resolve coherence scales and substructure \citep{Anderson2021, Anderson2024}, while all-sky Galactic and IGM studies have been driven by surveys at $\sim$1\,RM\,deg$^{-2}$ and are now advancing with POSSUM-like densities. Based on the faint polarised source population and the sensitivity needed for a meaningfully deeper grid, an SKA-Mid Band~2 polarisation survey is expected to reach $\sim$60--90 polarised sources per square degree \citep{galaxies8030053}, roughly a factor of two denser than current POSSUM-like depths, and to meet or exceed the required densities for cluster, Galactic, and large-scale structure science over large areas \citep{galaxies8030053, GaenslerPOSSUM}. In summary, the SKA represents a step change: denser and more precise RM grids, full-sky capability with complementary SKA-Low and SKA-Mid coverage, finer angular resolution, and improved spatial sampling and Faraday depth recovery across clusters, the Galactic ISM, and the cosmic web.

The combination of RM grids with other observational techniques will provide powerful synergies. For example, RM grids can be combined with X-ray observations of galaxy clusters to study the relationship between magnetic fields and thermal gas properties. RM grids can also be combined with observations of synchrotron emission to study the relationship between magnetic fields and relativistic particle populations. The SKA's multi-frequency capabilities will enable these combined analyses across a wide range of frequencies, providing comprehensive views of magnetic field structures and their role in various astrophysical processes.

\clearpage

\bibliographystyle{abbrvnat-maxbibnames4}
\bibliography{aaskaii_crossref,chapter}

\end{document}